\documentclass[fleqn,usenatbib]{mnras}

\usepackage{hyperref}
\usepackage{graphicx}	
\usepackage{amsmath}	
\usepackage{amssymb}	
\usepackage{multicol}        
\usepackage{bm}		
\usepackage{pdflscape}	

\newcommand{\Msun}{{\rm  M_{\odot}}}
\newcommand{\himpc}{{h^{-1}{\rm Mpc}}}
\newcommand{\hmpci}{{h{\rm Mpc}^{-1}}}
\newcommand{\himsun}{{h^{-1}{\Msun}}}
\newcommand{\temp}[1]{T_{\rm #1}}

\newcommand{\RA}[1]{{\color[gray]{0.} #1}} 
\newcommand{\RAm}[1]{{\color{black} #1}}

\newcommand{\HI}{\textsc{Hi}}
\newcommand{\PHIon}[1]{P_{\HI,{\rm on}}(#1)}

\newcommand{\PDMso}[1]{P_{\rm DM,so}(#1)}

\usepackage[T1]{fontenc}
\usepackage{ae,aecompl}

\usepackage{newtxtext,newtxmath}

\title[HI power spectrum reconstruction]{
Reconstructing {\HI} power spectrum with minimal parameters using the dark matter distribution beyond halos
}

\author[R. Ando et al.]{
Rika Ando$^{1}$\thanks{Contact e-mail: \href{mailto:ando.rika@b.mbox.nagoya-u.ac.jp}{ando.rika@b.mbox.nagoya-u.ac.jp}}
Atsushi J. Nishizawa$^{2,1}$,
Ikko Shimizu$^{3}$,
and 
Kentaro Nagamine$^{4,5,6}$
\\
$^{1}$Graduate School of Science, Nagoya University, Furocho, Chikusa Nagoya 461-8602, Aichi, Japan\\
$^{2}$Institute for Advanced Research, Nagoya University, Furocho, Chikusa Nagoya 461-8602, Aichi, Japan\\
$^{3}$Shikoku Gakuin University, 3-2-1 Bunkyocho, Zentsuji, Kagawa, 765-8505, Japan\\
$^{4}$Department of Earth and Space Science, Graduate School of Science, Osaka University, 1-1 Machikaneyama, Toyonaka, Osaka 560-0043, Japan\\
$^{5}$Kavli IPMU (WPI), The University of Tokyo, 5-1-5 Kashiwanoha, Kashiwa, Chiba, 277-8583, Japan \\
$^{6}$Department of Physics and Astronomy, University of Nevada, Las Vegas, 4505 S.Maryland Pkwy, Las Vegas, NV, 89154-4002, USA\\
}

\date{}

\pubyear{}

\begin{document}
\label{firstpage}
\pagerange{\pageref{firstpage}--\pageref{lastpage}}
\maketitle

\begin{abstract}
Intensity mapping of 21-cm line by several radio telescope experiments 
will probe the large-scale structure of the Universe in the post-reionization epoch. 
It requires a theoretical framework of neutral hydrogen ({\HI}) clustering, such as modelling of {\HI} power spectrum for Baryon Acoustic Oscillations (BAO) analysis.
We propose a new method for reconstructing the {\HI} map from dark matter distribution using N-body simulations. 
Several studies attempt to compute the {\HI} power spectrum with N-body simulations by pasting {\HI} gas at the dark matter halo centre, assuming the relation between the halo and {\HI} masses. 
On the other hand, the method proposed in this paper reproduces the {\HI} power spectrum from simulated dark matter distribution truncated at specific scales from the halo centre. 
With this method, the slope of {\HI} power spectrum is reproduced well at the BAO scales, $k<1 h/{\rm Mpc}$. 
Furthermore, we find the fluctuation of spin temperature, which is often ignored at the post-reionization epoch, alters the power spectrum of brightness temperature by at most 8\% in the power spectrum.
Finally, we discuss how our method works by comparing the density profiles of {\HI} and dark matter around the dark matter halos. 
\end{abstract}




\begingroup
\let\clearpage\relax
\endgroup
\newpage

\section{Introduction}
\label{sec:introduction}
Recently, 21-cm line signal is drawing much attention in the context of both cosmology and astrophysics \citep{Furlanetto+2006}. The 21-cm line is a radio wave emitted from the neutral hydrogen ({\HI}) due to the hyperfine splitting, and its rest-frame frequency is 1420 MHz. 
\RAm{The 21-cm line IM is a method to measure
the line intensity without resolving individual galaxies. 
Therefore, IM enables us to cover a wide area of sky efficiently. Aside from the interlopers, the redshift determination accuracy simply depends on the frequency resolution of the instruments.
Currently, some experiments of 21-cm intensity mapping (IM) are planned and under construction, such as the 
TianLai~\footnote{http://tianlai.bao.ac.cn/index.html} \citep{Chen2012},
Baryon acoustic oscillations In Neutral Gas Observation (BINGO)~\footnote{https://www.bingotelescope.org/en/bingo/} \citep{Battye+2012}, 
Canadian Hydrogen Intensity Mapping Experiment (CHIME)~\footnote{https://chime-experiment.ca/en} \citep{Newburgh+2014}, 
Hydrogen Intensity and Real-time Analysis eXperiment (HIRAX)~\footnote{https://hirax.ukzn.ac.za/} \citep{Newburgh+2016}
and 
Square Kilometre Array (SKA)~\footnote{https://www.skatelescope.org/} \citep{Santos+2015}.}

At the post-reionization epoch, almost all hydrogen atoms are ionized, and {\HI} resides only in high-density regions. 
\RAm{Therefore, {\HI} can be a good tracer of the matter density distribution, and the 21-cm line IM is a new probe of the large-scale structure \citep{Bull+2015} at higher redshifts, such as $z>3$. Quasar observations and other galaxy surveys such as DESI, PFS or HETDEX can reach those redshifts, and 21-cm line IM will be a comprehensive probe.}

\RAm{The BAO, which is a feature imprinted on the matter density field with a characteristic scale ($\sim 100\himpc$),
is recognized as the most promising and robust method for constraining dark energy through the Hubble parameter and the angular diameter distance  \citep{Eisenstein+2005}. 
}
\RAm{Recent observations of galaxy redshift surveys enable us to measure the BAO with various tracers such as luminous red galaxies \citep{GilMarin+:2020}, quasi-stellar objects (QSOs) \citep{Ata+2018,Zarrouk+2020} or Lyman-$\alpha$ forest \citep{eBOSSLya:2020}, and have shown that the BAO scale is determined within a few per cent accuracy.} 
Therefore, the required accuracy of the BAO scale determination in the next decade must be 1\% or better. With the 21-cm line IM, we can extend these analyses to the higher redshifts, but at the same time, the accurate prediction of the BAO scale for 21-cm line IM is desired.

Precise modelling of {\HI} is useful not only for cosmology but also for astrophysical context.
For example, the density profile of {\HI} around active galactic nuclei (AGN) is related to the galaxy properties and its environment \citep{Mukae+2020,Momose+2020}. 
These studies use IGM tomography by Lyman-$\alpha$ forest. 
It measures the amount of {\HI} along the line of sight by observing the subsequent absorption lines imprinted on the spectrum of a background object. 
Previously, QSOs have long been the most promising target as backlight sources
because of their brightness \citep{Croft+1998,Busca+2013}. 
However, the angular resolution is limited by the surface number density of background sources, and the sightline density of QSOs is not so high. 
\cite{Lee+2014b} proposed to use the Lyman-break galaxies (LBGs) as backlight sources 
since the number density of faint LBGs \citep[e.g., $g > 24$\,mag;][]{Reddy+2008} is much higher than that of QSOs \citep{Palanque-Delabrouille+2013}.
The Hyper-Suprime Cam survey has detected many LBGs \citep{Ono+2018}, and their spectra are going to be taken by
the Prime Focus Spectrograph (PFS) installed on the Subaru telescope. It will observe thousands of galaxies in a single field of view \citep{PFS2014}, and it enables us to measure the 3-dimensional {\HI} distribution more precisely \citep[see the appendix of][]{Nagamine20}.

\RAm{The distribution of {\HI} depends not only on the gravitational force but also on the highly complicated astrophysical processes such as non-trivial UV background radiation or the feedback from supernovae or AGNs. }
Therefore, it is difficult to predict the {\HI} power spectrum based solely on the physically motivated analytic models. 
Previously, researchers used cosmological hydrodynamic simulations to measure the large-scale {\HI} clustering \citep{TNG2018, Ando+2019}. 
However, these simulations' box sizes are insufficient to measure the power spectrum at the BAO scales, $k\simeq 1h$\,Mpc$^{-1}$. 
It is always challenging to achieve high mass resolution in a large simulation box size due to computational cost.
\RAm{If we insist on the large-scale simulation sacrificing the mass resolution, we may lose a non-negligible
amount of {\HI}, which resides in the low-mass halos, and thus 
the shape and amplitude of the {\HI} power spectrum depends on the resolution of the simulation \citep[see Appendix of][]{TNG2018}.} 
This is a sort of artefact which we should avoid when we model the {\HI} power spectrum.
One of the possible ways to model the {\HI} distribution is 
to use the simulated dark matter halo distribution with the relation of $M_{\HI}-M_h$ calibrated a priori using the results of high-resolution cosmological hydrodynamic simulations 
\citep{Sarkar+2016,Sarkar+2018,Modi+:2019,Wang+:2019}. 
This method seems reasonable since most of the {\HI} is localized within dark matter halos. 
Although the method only provides approximate {\HI} distribution, it allows us to reduce the computational cost significantly. However, we note, that the method depends on the $M_{\HI}-M_h$ relation, which is highly dependent on the assumed baryonic physics and the mass resolution of the hydrodynamic simulation. 
Instead of using the simulated distribution of dark matter, focusing on predicting the power spectrum, \cite{Padmanabhan+2017a,Padmanabhan+2017b} have applied an analytic halo model empirically assuming the {\HI} profile within the halo. 

Those previous works implicitly assume that all the {\HI} is confined within the dark matter halo. However, in this paper, we first demonstrate that a certain fraction of the {\HI} resides not only within the halo but also extends to the outside of halos, i.e., inter-galactic medium (IGM). 
\RAm{Hence in this paper, we investigate the impact of the \HI\ in the IGM to the power spectrum as a function of the halo centric radius. Based on our experiments, we introduce a new method to reconstruct the {\HI} power spectrum from the dark matter only N-body simulation. This method does not rely on the $M_{\HI}-M_h$ relation which is computationally expensive to calibrate, and only based on the fact that the {\HI} is localized within and around the dark matter halo.
}

The paper is organised as follows. 
In Sec.\ref{sec:simulation}, we first describe 
the IllustrisTNG simulation and how to construct the density fields. 
In Sec. \ref{sec:HImap}, we present the method of reconstruction of the {\HI} density field and the main result.
In Sec.\ref{sec:temperature}, we show a detailed treatment of the 21-cm signal, the comparison with the previous method, and the interpretation using the density profile. 
Finally, Sec.\ref{sec:summary} summarises the paper. 
Throughout this paper, we adopt the Planck 2015 cosmological parameters \citep{Planck15}, which is also adopted in the IllustrisTNG simulations. 
\RAm{
In this paper, we focus on the analysis at redshift $z=3$, where the up coming radio interferometer, SKA can probe.
}

\section{Simulation}
\label{sec:simulation}
To investigate the 
spatial distribution of {\HI} after reionization, and to explore a method to reproduce the {\HI}
power spectrum from the dark matter distribution, we mainly
use 
the IllustrisTNG simulations. 
In this section, we briefly revisit the IllustrisTNG simulation and describe how we construct the {\HI} gas density and brightness temperature distributions from the simulation. 

\subsection{Illustris TNG simulation}
The IllustrisTNG  \citep{Nelson+2019,Nelson+2018,Marinacci+2018,Springel+2018,Pillepich+2018,Naiman+2018} is cosmological magnetohydrodynamical simulation series 
based on the original Illustris simulations \citep{Nelson+2015,Vogelsberger+2014a,Vogelsberger+2014b,Genel+2014,Sijacki+2015}. 
The IllustrisTNG solves magnetohydrodynamics with the moving-mesh code \texttt{AREPO} \citep{Springel+2010}.
TNG100 and TNG300 simulations are publicly available
with a box size of $75\himpc$ and $205\himpc$ on a side, respectively. 
Each realization has three different mass resolutions. 
It has been shown that the {\HI} power spectrum is highly deformed by the lack of mass resolution of the simulation \citep{Ando+2019}. 
This is simply because of the assumption that the {\HI} resides only in the halo, and if we have a low mass resolution, smaller halos do not form in the simulation. Then the bias of {\HI} power spectrum becomes higher than the one from higher-resolution simulations. 
In order to build a method for reconstructing the {\HI} map that reproduces the realistic {\HI} power spectrum, we use TNG100-1, which has the highest resolution with $2\times 1820^3$ dark-matter (DM) particles and gaseous cells. 
The mass resolution is $5.1\times 10^6 \himsun$ for dark matter particle and $9.4\times 10^5 \himsun$ for the gaseous cell. 
The minimum value of the gravitational softening in comoving units is $\epsilon_{\rm DM,min}=0.5h^{-1}\,{\rm kpc}$ for dark matter particles 
and $\epsilon_{\rm gas,min}=0.125h^{-1}\,{\rm kpc}$ for gas.
The IllustrisTNG simulations include physical models for star formation and feedback by supernovae and active galactic nuclei. 

The ionization rate of hydrogen is calculated on-the-fly and both photo-ionization and collisional-ionization are taken into account.
The spatially uniform and redshift-dependent UV background (UVB) radiation \citep{FG11} is adopted at $z<6$ and self-shielding correction is included in the simulation. 
A part of the gas particles can be transformed into stars or black holes (BH), and the energy released from the supernova and super-massive black hole are considered. 
\RAm{Two different AGN feedback modes were considered in the IllustrisTNG simulation: the quasar mode (distributing thermal energy like quasars at high accretion rate) and the radio model (kinetic wind from AGN radio jets at low accretion rate). }

Several galaxy formation models in the IllustrisTNG are updated from the original Illustris simulation. 
Here we only describe the two major updates but refer the reader to \cite{Pillepich+2018a} for more complete descriptions.
First, the low-accretion rate mode for AGN feedback uses kinetic wind \citep{Weinberger+2017} instead of the bubble model \citep{Sijacki+2007}. 
This modification improves the discrepancy between the Illustris and the observations, such as the galaxy stellar mass function on the massive side. 
It also increases the overall HI fraction, in particular at the halo region.
Second, the magnetic field is introduced, and this affects the massive halo and suppress the star formation. 
\cite{Pillepich+2018} shows that these different treatments for the astrophysical effects may bring prominent effects on the temperature or {\HI} density field on cosmological scales.

Another important piece of data for our analysis is dark matter \textit{halo}.
The IllustrisTNG database provides the halo catalogue based on the friend-of-friend (FoF) algorithm with a linking length parameter $b=0.2$. 
We use the $R_{200c}$ (\texttt{Group\_R\_Crit200}, the radius where the mean density inside is 200 times the critical density) and 
$M_h$ (\texttt{GroupMass}, sum of all particles in each halo) to define the virial radius and mass of halo, and we assign the halo at position of minimum gravitational potential, \texttt{GroupPos}. 
Other papers used the centre of mass, \texttt{GroupCM}, as the definition of halo position. 
We note that \texttt{GroupPos} and \texttt{GroupCM} differ significantly; For some halos, the offset between them is comparable or even larger than the virial radius. It is not obvious which halo centre is better to use; however, we see that the {\HI} is more concentrated around the \texttt{GroupPos} and we take this as the centre of the halo in this paper. 

\subsection{Calculation of density fields}
\label{ssec:densityfield}
In this section, we describe how to measure the spatial inhomogeneity of the 21-cm intensity from the simulation. As shown later in this section, the sources of inhomogeneity arise from gas density and neutral fraction fluctuations and the spatial variation of spin temperature of neutral hydrogen, where the latter is often ignored at the epoch of post reionization in the literature. \RAm{We discussed the {\HI} clustering in real space, and the HI power spectrum in redshift space is discussed in \cite{TNG2018}.}

We use 
the Nearest Grid Point (NGP) method to construct the {\HI} and dark matter density fields from the particle distribution. 
The fluid element in the IllustrisTNG simulation occupies a voronoi tessellation cell. 
However, the size of the voronoi cell is negligibly small compared to our grid size if we take $N_{\rm grid}=1024^3$, and thus we can treat gaseous cells as point-like object just like the dark matter particles. 
\RAm{With the mass density fluctuation $\delta = \rho/\bar{\rho}-1$, we compute the power spectrum defined as 
\begin{equation}
	\label{eq:def_powerspectrum}
	P_{\rm \alpha}(k_i) = \frac{1}{N_{k_i}} \sum_{\{j; k_j \in k_i\}} \delta_\alpha(\boldsymbol{k}_j) \delta_\alpha^*(\boldsymbol{k}_j),
\end{equation}
where the subscript $\alpha$ is one of {\HI}, DM or $\delta T_b$. } 
$N_k$ is the number of the modes available in each bin of wavenumber.
We approximate the variance of the power spectrum as \citep{Bernardeau+2002}
\begin{equation}
    \sigma^2 \left( P_{\alpha}(k_i) \right) \approx \frac{2P_{\alpha}^2(k_i)}{N_{k_i}},
\end{equation}
in the cosmic variance-limited regime.

In a practical observation, we will see the 21-cm line signal as the difference of the temperature between the source and the background radiation, and observe it as emission or absorption lines.
The intensity of radio wave we observe can be translated to the brightness temperature as \citep[e.g.,][]{Field1958}
\begin{equation}
    \delta \temp{b} = \frac{\temp{S}-\temp{\gamma}(z)}{1+z}\left(1-e^{-\tau_{\nu_0}}\right),
\end{equation}
where $\nu_0=1420.4057 {\rm MHz}$ is the rest frame frequency of the 21-cm line, $\tau_{\nu_0}$ is the optical depth of the {\HI} cloud, $\temp{S}$ is the spin temperature of the neutral hydrogen, and $\temp{\gamma}(z)$ is CMB temperature as the backlight.
\RAm{The optical depth which is the absorption coefficient integrated over the path-length can be expressed in the following formula,
\begin{equation}
    \tau_{\nu_0} = \frac{3}{32\pi}\frac{hc^3A_{10}}{k_{\rm B}\temp{S}\nu_0^2}\frac{n_{\HI}}{(1+z)(dv_{\parallel}/dr_{\parallel})},
\end{equation}
where 
$n_{\HI}$ is the \HI\ number density, and 
$dv_{\parallel}/dr_{\parallel}$ is the velocity gradient of the source along the line of sight, which is the sum of the Hubble parameter $H(z)$ and the gradient of the peculiar velocity ($dv_{\parallel}/dr_{\parallel}=H(z)+\Delta H$). 
This is induced by the redistribution of the emission due to the Doppler effect and blurs the line profile. The broadening effect associated with the peculiar velocity, $\Delta H$
is widely thought to be small compared to the Hubble expansion and from the simulation we see that $\langle \Delta H/H\rangle = 1.0\times 10^{-5} \,^{+0.3}_{-0.04}$. Therefore, we omit the effect of peculiar velocity and replace $dv_{\parallel}/dr_{\parallel}\simeq H(z)$.} 
Since the optical depth is small enough, the differential brightness temperature can be rewritten as 
\begin{align}
    \delta T_b &\approx 
     \frac{3}{32\pi}\frac{hc^3A_{10}}{k_{\rm B}\nu_0^2}\left(1-\frac{T_{\gamma}(z)}{\temp{S}}\right)\frac{n_{\HI}}{(1+z)H(z)}.
\end{align}
If we ignore the peculiar velocity, the \HI\ density and the spin temperature only fluctuate spatially in the differential brightness temperature. 

The spin temperature $\temp{S}$ is determined by three physical processes, and is expressed by using the temperatures that characterise each of them as follows \citep{Field1958}:
\begin{equation}
    \temp{S}^{-1} = \frac{\temp{\gamma}^{-1}+x_{\alpha}\temp{\alpha}^{-1}+x_c\temp{K}^{-1}}{1+x_{\alpha}+x_c}
\end{equation}
where $\temp{\alpha}$ and $x_{\alpha}$ are the colour temperature of Ly$\alpha$ radiation field and its coupling coefficient, and $\temp{K}$ and $x_c$ are the kinetic gas temperature and its coupling coefficient.
The Wouthuysen-Field (WF) effect is the effect that Ly$\alpha$ photons excite hydrogen once and then cause spin-flips when 
deexcited \citep{Wouthuysen1952, Field1958}. The colour temperature is approximated to the gas kinetic temperature $\temp{c}\simeq \temp{K}$ due to the atomic recoil.

At the epoch of the post reionization, the spin temperature is coupled to the gas temperature, and the gas temperature is much higher than the CMB temperature. 
Therefore, we can approximate $1-\temp{\gamma}/\temp{S} \rightarrow 1$ and the spatial fluctuation caused by $\temp{S}$ disappears from the $\delta \temp{b}$. Now the power spectrum of $\delta \temp{b}$ is simply proportional to that of {\HI}, 
\begin{equation}
    P_{\delta T_b}(k) \propto P_{\HI}(k). 
\end{equation}
Although we drop the spin temperature-dependent term in our fiducial analysis, we see that the inhomogeneous spin temperature imprints several per cent fluctuations on the power spectrum (see Section \ref{ssec:spintemp} for more details). 

\subsection{Corrections for star-forming gas particles}
\label{ssec:stargas_method}
In Sec\ref{ssec:densityfield}, we 
estimate {\HI}, directly from the output 
of IllustrisTNG 
using {\HI} fraction and mass of the gas particles. 
However, it is pointed out that the prescription of \cite{Springel+2003} underestimates the total amount of {\HI} 
compared to the damped Lyman-$\alpha$ observations \citep{Rao+2006, Lah+2007,Noterdaeme+2012,Songaila+2010,Crighton+2015}.
Instead, \cite{TNG2018} recompute the {\HI} mass by solving the equilibrium equation by fixing the temperature for the star-forming gas to $10^4$ K \citep{Rahmati+2013}. This greatly increases the {\HI} mass, especially for the massive halos. We see that the effect can be separated into two distinct effects; the intrinsic increase of the UV photon, and the effect of self-shielding. Both effects increase the mass of {\HI}; e.g., for $10^{11} {\rm M_\odot}$ halos, the {\HI} mass increases by a factor of a few than those without correction.
Another correction has also been made 
to account for the $\rm H_2$ abundance. 
This hardly decreases the {\HI} mass, and it can be safely ignored. 
In our fiducial analysis, these corrections have not been applied when calculating the {\HI} mass (Section \ref{sec:HImap}), but we will see the effect of the corrections to our results in Section~\ref{ssec:stargas}.

\section{Reconstructing \HI{} power spectrum from DM simulation}
\label{sec:HImap}

We first examine the \HI\ distribution in Sec~\ref{ssec:hiso}, and develop a new method to reconstruct the {\HI} map from dark matter distribution in Sec~\ref{ssec:dmso}. 
In this section, the \HI\ density field is calculated using the output data from IllustrisTNG without the star-forming gas correction.

\subsection{SO-HI halo}
\label{ssec:hiso}
\begin{figure}
\centering
	\includegraphics[width=\linewidth]{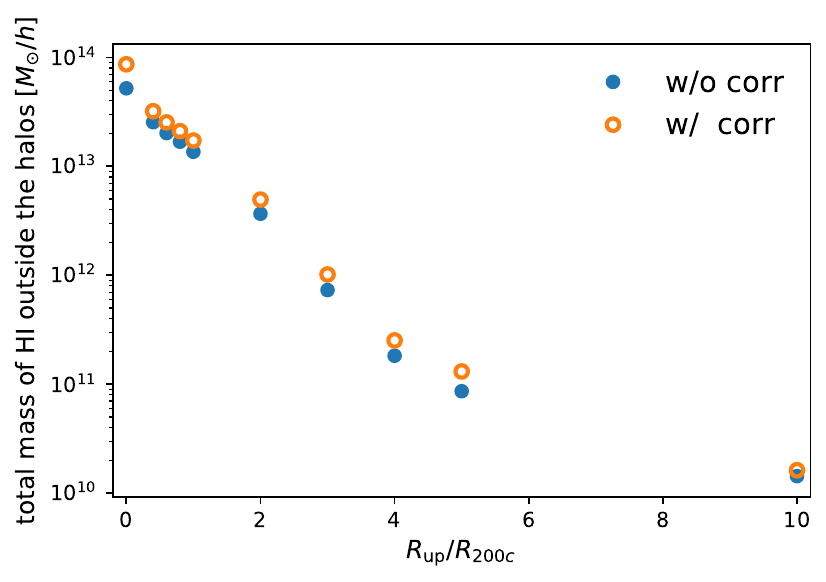}
    \caption{
    Total \HI\ mass outside the radius from the centre of the halo as a function of the ratio $R_{\rm up}/R_{200c}$. The value at $R_{\rm up}/R_{200c} = 0$ is equivalent to the total {\HI} mass in the simulation box. 
	    \label{fig:HIratio}}
\end{figure}

For the precise modelling of the {\HI} power spectrum, it might be helpful to understand where the {\HI} is populated. 
In the epoch of post-reionisation, most of the {\HI}
reside inside dark matter halos. 
In order to quantify this, we compute the mass of {\HI} as a function of halo-centric distance. 
Figure~\ref{fig:HIratio} shows the total {\HI} mass that is located at outside of the halo, $R>R_{\rm up}$. In this paper, we often call the region where the halos are included as \textit{spherical overdensity} (SO) region, with a specific radial scale $R_{\rm up}$ which plays a central role in our model. 
From this analysis, we find that 
$74\%$ of total {\HI} mass is confined within $R_{\rm up}=R_{200c}$ and $93\%$ within $R_{\rm up}=2R_{200c}$. 
Compared to this, only 16\% of total dark matter is confined within the halo resolved in the same simulation. This implies that {\HI} is extremely localised to the halo interior. 

\begin{figure}
\centering
	\includegraphics[width=\linewidth]{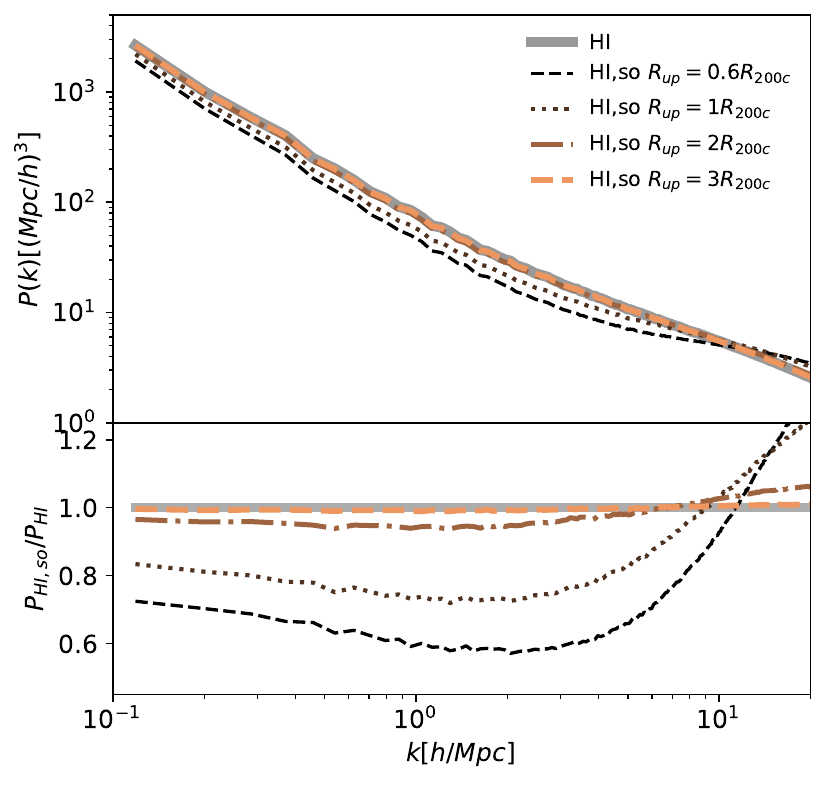}
    \caption{
    The power spectrum of {\HI} within different halo-centric radii.
    The solid thick grey line shows the power spectrum using all \HI\ included in the simulation box, whilst the other dashed lines are from {\HI} in the vicinity of halos with radii shown in the legend. Bottom panel shows the fractional difference from total {\HI} power spectrum.
	    \label{fig:HImassratioSO}}
\end{figure}

For completeness, we compare the power spectrum of {\HI} derived from the entire simulation box and the one from {\HI} 
at $R<R_{\rm up}$.
As can be seen from Fig.~\ref{fig:HImassratioSO}, the difference in power spectrum is significantly reduced as we increase the radius $R_{\rm up}$, and we find that the power difference is less than 1\% if we take $R_{\rm up}=3R_{200c}$. 
This implies that we need to understand the distribution of {\HI} within the halo and around the halo within $R<3R_{\rm 200c}$.

\subsection{SO-DM halo}
\label{ssec:dmso}
In this section, we consider reconstructing the {\HI} distribution from only dark matter simulations. If we model {\HI} power spectrum only using dark matter, this greatly reduces the computational cost. 
Sometimes in the literature, the {\HI} gas is pasted onto dark matter halo using
the relationship between $M_{\rm halo}$ and $M_{\HI}$, which should be calibrated using hydrodynamic simulations \citep{TNG2018, Sarkar+2016, Sarkar+2018}.
However, in fact, not only the {\HI} gas within the halos but also those in the vicinity of halos contribute to the {\HI} power spectrum. 
Furthermore, 
this method highly depends on the assumed $M_{\rm halo}-M_{\HI}$ relation, 
which is measured from hydrodynamic simulation or obtained by using a model with observational data \citep[e.g][]{Wang+:2019}.
Therefore, we propose a more general method for reconstructing the $P_{\HI}$ which does not rely on the $M_{\rm halo}-M_{\HI}$ relationship.

\begin{figure*}

    \begin{tabular}{cc}

    \centering
	\includegraphics[width=0.45\linewidth]{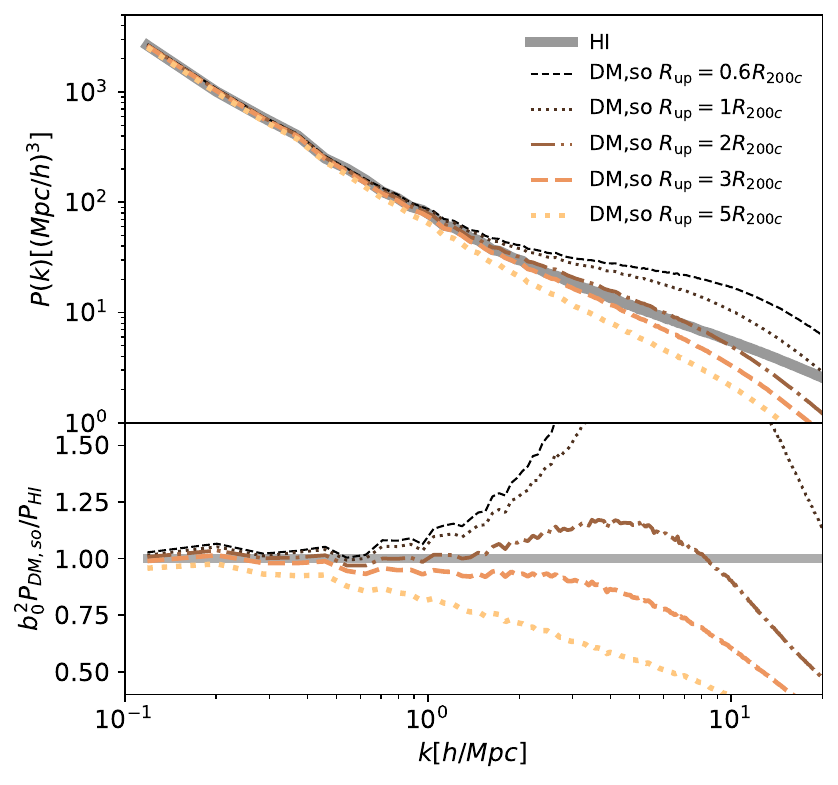}&

\centering
 	\includegraphics[width=0.45\linewidth]{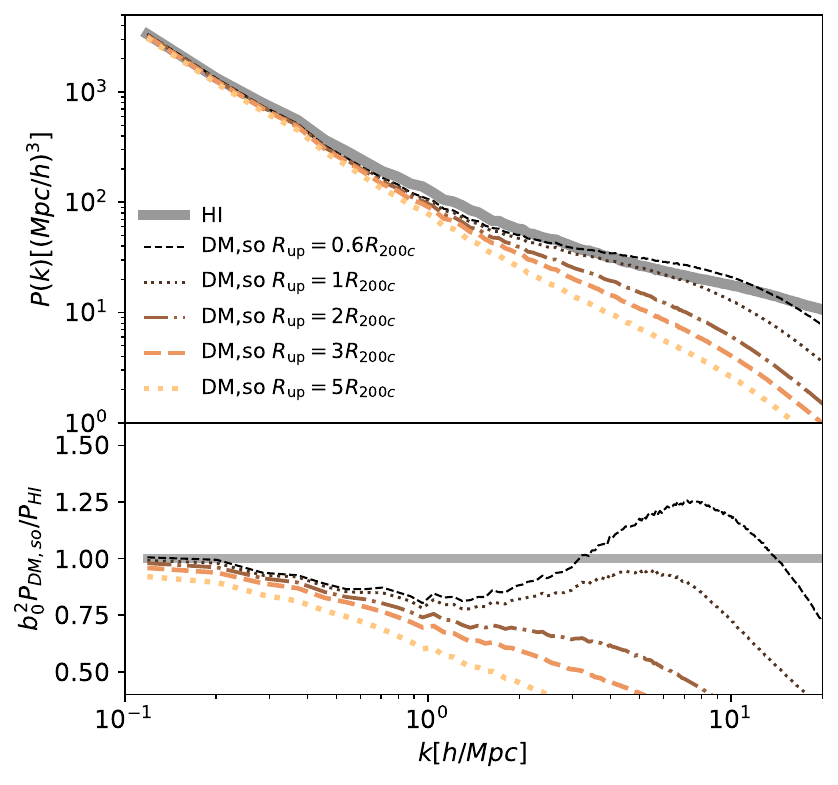}	

	    \end{tabular}
    \caption{
    \textit{Left panels:} Upper panel shows $b_0^2\,P_{\rm DM,so}(k)$ and the {\HI} power spectrum without the star-forming gas correction. 
    Bottom panel shows the ratio of the reconstructed power spectra to the true {\HI} power spectrum. 
    \textit{Right panels:} Same as the left panels, but with the star-forming gas correction. 
	    \label{fig:Pdmso_off}}
\end{figure*}

We focus on the dark matter distribution around halos. Using only dark matter N-body simulation, we compute the dark matter density field using the particles around halos. 
Then we compute the power spectrum for this biased sample of dark matter, $P_{\rm DM,so}(k)$. This biased selection effectively corresponds to applying the halo-like bias to the dark matter power spectrum (i.e. If we take $R_{\rm up}$ equal to the virial radius, this exactly corresponds to the halo bias.). 
The simplest case is that the {\HI} is linearly biased to dark matter, but given that the {\HI} is highly localised around the halos, it is a reasonable expectation that the {\HI} is linearly biased to dark matter around halos.  \RAm{Then our model leads
\begin{equation}
\delta_{\rm DM,so}({\bm x}) = 
\delta_{\rm DM} ({\bm x})  
\,\times\,\Theta\left(
\sum_i 
W^{\rm TH}_{R_{\rm up}}(|{\bm x}-{\bm x}_i|) 
\right),
\end{equation}
where index $i$ runs over all halos, $x_i$ denotes the central position of halo, $\Theta$ is a step function, $\Theta(X)=1$ for $X\geq1$, and 0 otherwise, and $W^{\rm TH}_{R_{\rm up}}$ is a top-hat window function. 
In this way, we generate a pseudo-scale dependent halo-biased density field by extracting dark matter particles only in and around the halo. 
As a result, the slope of the power spectrum of this density field becomes close to that of the \HI\ power spectrum. 
For the BAO analysis, the constant amplitude offset of the power spectrum is not important and we only focus on the scale dependence of the 
model spectrum. We define the model residual as
${\rm Res}(k)=\sqrt{P_{\HI}/P_{\rm DM,so}}$
to quantify how our model reproduces the scale dependence of the {\HI} power spectrum. 
In order to isolate the constant offset, we describe the model residual as
\begin{equation}
    \label{eq:kfitfunc}
    {\rm Res}(k) = b_0+b_1 k,
\end{equation}
and see if the value of $b_1$ is consistent with zero. 

The left panel of Fig.~\ref{fig:Pdmso_off} 
compares the reconstructed \HI\ power spectra from dark matter-only N-body simulation with 
different $R_{\rm up}$ 
with total \HI\ power spectrum. 
We find that the reconstructed power spectrum with $R_{\rm up}=2R_{200c}$ best reproduces the true \HI\ power spectrum. Therefore, we can reconstruct the \HI\ power spectrum $P_{\HI}(k)$ on large scales
by using dark matter density field inside $R_{\rm up}\leq 2R_{\rm 200c}$ from the halo centre. 
This result is valid only at $z=3$, and at other redshifts, we find different values   of $R_{\rm up}$ that is most appropriate. This will be discussed in Section \ref{ssec:assessing}.}

\section{Discussion}
\label{sec:temperature}
\subsection{Correction for star-forming particles}
\label{ssec:stargas}
In Sec~\ref{ssec:densityfield}, we described the results obtained from the \HI\ data in IllustrisTNG simulation. 
In this section, we use the \HI\ density where \HI\ fraction of the star-forming gas 
is corrected as described in Sec~\ref{ssec:stargas_method}. 
This correction increases the {\HI} fraction of the star-forming gas and therefore has the effect of increasing the {\HI} mass in dense regions. 
Especially, the effect of this correction is more significant 
for massive halos, which means that the {\HI} mass in massive halos increases. 
This makes the {\HI} bias larger and we see that the amplitude of $P_{\HI}$ increases with this correction. We further explore the picture with the help of a halo model \citep{Seljak+2000}. 
\RA{The halo bias is in general stronger for more massive halos, and the star-forming correction increases the {\HI} gas in particular in massive halos. Therefore, the correction effectively increases the {\HI} bias and the amplitude of the {\HI} power spectrum is boosted on large scales.
Also, it is well known that the 2-halo term has a truncation from the linear power spectrum depending on the effective scale of halo radius. 
This truncation scale is shifted to a larger scale. \RAm{This is because the star-forming gas correction selectively increases 
the {\HI} mass for the massive halos and thus put more weights on the massive halos. Therefore the effective radius after correction is increased.}
On the other hand, if we focus on the smaller scales, the contribution of 1-halo term becomes more significant. Given that the massive halos have larger shot-noise (due to their small number), again the correction makes the relative contribution from the massive halos more prominent, and thus the amplitude of the 1-halo term also becomes larger.
While the increase of the 1-halo term on small scales makes the power spectrum flatter, the shift of the damping scale of the 2-halo term makes it steeper. Therefore, the resulting slope of the power spectrum is determined by the balance of these two competing effects. 
} 
If we compare the {\HI} power spectrum in Fig.~\ref{fig:Pdmso_off}, we see that the slope of $P_{\HI}(k)$ becomes shallower when including the star-formation correction. 

In the rest of this section, 
we examine if our prescription to describe the \HI\ power spectrum (Sec. \ref{sec:HImap}) is still valid even after making such a correction. 
We first expect that the best SO radius for reproducing the slope of the {\HI} power spectrum with star-forming gas correction, $\PHIon{k}$, is shallower than that for without the correction.  
In order to see this, let us again consider the picture in the context of the halo model. 
Decreasing the SO radius 
reduces the mass contained within the halo and thus it shifts the \HI\ mass function to the lower mass side. 
Since the mass function is monotonically decreasing function with mass, 
it means that the number of massive halos decreases at a given mass. 
This makes the shot noise of massive halos larger.
We note that it also increases the number of halos at the lower side, but since the number is so large that it is unlikely to contribute to the total shot noise.
At the same time, since the contribution of the small halos becomes relatively larger when $\tilde{\rho}(k,M)\rightarrow 0$, 
the damping scale of the 2-halo term shifts to the smaller scales. 
\RAm{Indeed, in Fig.~\ref{fig:Pdmso_off}, we see that the smaller SO radius $R_{\rm up}=0.6 R_{\rm 200c}$ is the most acceptable to reproduce the $\PHIon{k}$. 
However, the effect of changing the slope is marginal, and we do not see any difference between $\PDMso{k|R<0.6R_{\rm 200c}}$ and $\PDMso{k|R=0.6R_{\rm 200c}}$ and it best describes the slope of $\PHIon{k}$. Therefore, we adopt $R_{\rm up}=0.6R_{\rm 200c}$ to reproduce the 
$\PHIon{k}$. }

\subsection{Comparison to previous work: $M_{\HI}$ pasting}
\label{ssec:pasting}
In this section, we compare our results for $P_{\HI}(k)$ with the previous works \citep{TNG2018}. 
Some previous studies have placed the {\HI} mass at the centre of the dark matter halo to create the {\HI} map.
This procedure is sometimes called as `pasting', and we need to assume the relation between the mass of {\HI} and the dark matter mass of the host halo ($M_{\rm halo}-M_{\HI}$), which can be calibrated 
using hydrodynamic simulations, semi-analytical models, and observational data \citep{Sarkar+2016,Sarkar+2018,Wang+:2019,Modi+:2019}. 
\cite{TNG2018} fitted this relationship measured from the TNG100-1 run by the following function,
\begin{equation}
    M_{\HI}(M,z)=M_0\left(\frac{M}{M_{\rm min}}\right)^{\alpha}\exp\left[-\left(\frac{M_{\rm min}}{M}\right)^{0.35}\right]
    \label{eq:MMrelation}
\end{equation}
where $\alpha,M_0,M_{\rm min}$ are free parameters. $M_{\rm min}$ means the cut-off of the halo mass where {\HI} can resides. In that way, they reproduced {\HI} map, and  compared the {\HI} power spectrum measured from the hydrodynamic simulation with that reconstructed from the N-body simulation using this method.
We note that the distribution of $M_{\HI}$ is highly scattered around the above equation; however, including the scatter when pasting the {\HI} to the dark matter halos does not alter the resulting power spectrum \citep[e.g.][]{Modi+:2019}.

In order to compare 
the pasting method with our method, 
we follow the procedures in \citep{TNG2018}. For each halo, we sum up the {\HI} mass over all FoF gas particles included in that halo. When we calculate the {\HI} mass for each gas particle, we also consider the star-forming gas corrections described in Section~\ref{ssec:stargas}. The best-fitting parameters are summarised in Table \ref{tab:fitting} for both with and without correction.

\begin{figure*}
    \begin{tabular}{cc}
\centering
	\includegraphics[width=0.45\linewidth]{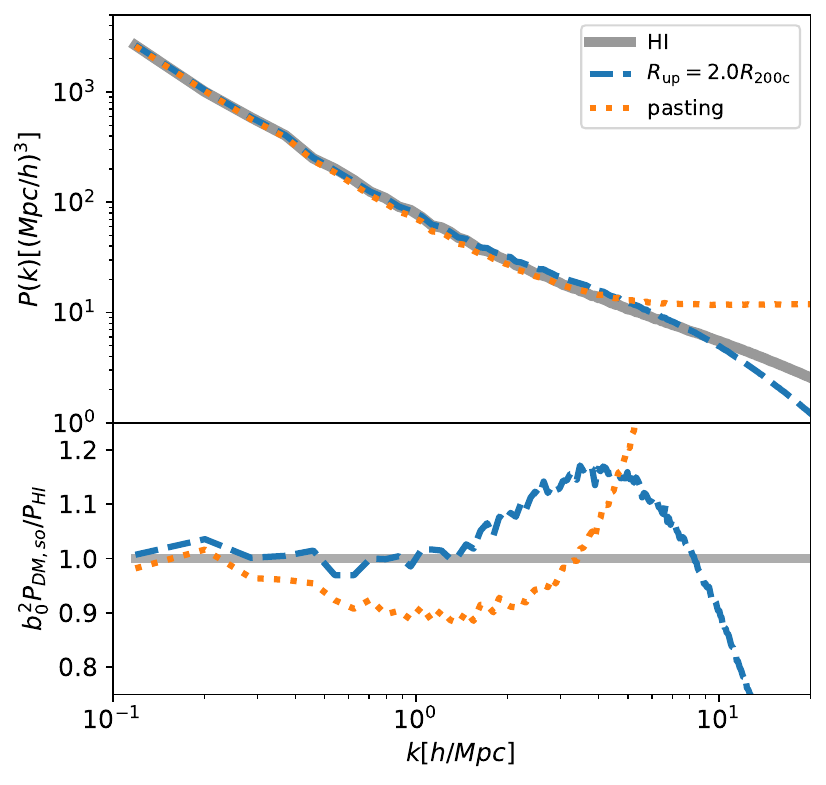}&
\centering
	\includegraphics[width=0.45\linewidth]{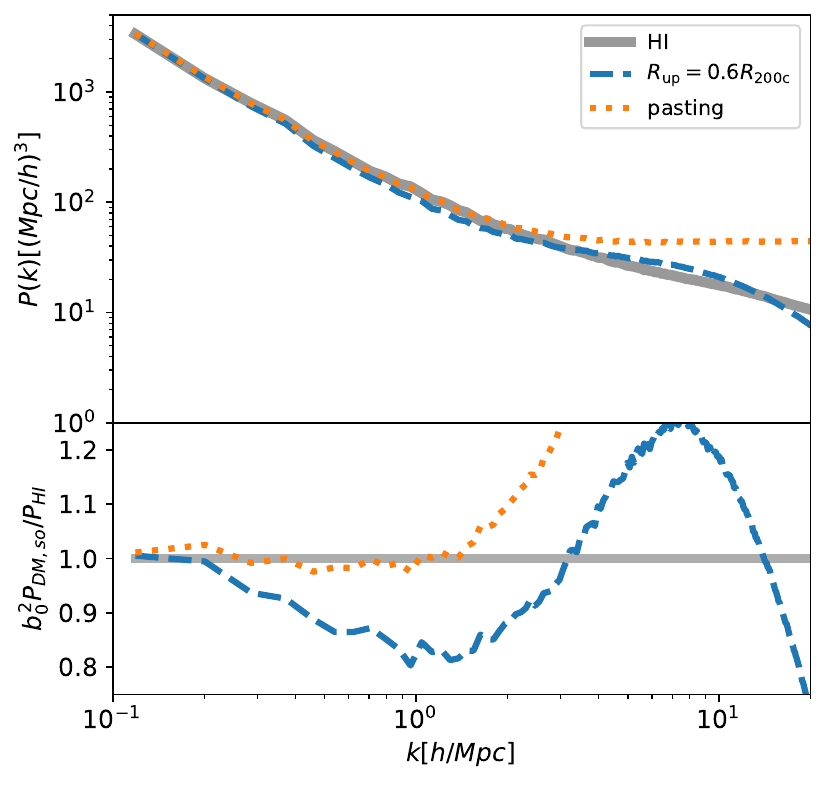}
	        \end{tabular}
    \caption{
    {\it Left panels:} Upper panel shows the true {\HI} power spectrum measured in the Illustris hydrodynamic simulation without star-forming gas correction $P_{\HI,\rm off}(k)$ (grey line), and those reconstructed from N-body simulation using $M_{\rm halo}-M_{\HI}$ method (orange-dotted line) and the SO method (blue-dashed line). Bottom panel shows the ratio between the reconstructed power spectrum and the true {\HI} power spectrum.
    {\it Right panels:} Same as the left panel, but with the star-forming gas correction.
    }
	    \label{fig:Precon_off}

\end{figure*}

The result is shown in Fig.~\ref{fig:Precon_off} for the case without (left panel) and with (right panel) the star-forming gas correction, respectively.
Like in the BAO analysis, if we focus on the slope of the power spectrum, we see that our method reproduces the slope of {\HI} power spectrum $P_{\HI,\rm off}$ better than the {\HI} pasting in the case without the correction. On the other hand, once the star-forming gas correction is applied, the {\HI} pasting better represents the {\HI} power spectrum than our best-describing model with $R_{\rm up}=0.6R_{\rm 200c}$.

The goodness of the model can be recognised by how well we can reproduce the slope of the power spectrum of {\HI} at the scale of interest, $k<1 h/{\rm Mpc}$. 
\RAm{Without the correction, $b_1 = 0.07 \pm 0.03$  for the pasting, and $b_1 = 0.005 \pm 0.03$ for the SO method with the radius $R_{\rm up}=2R_{200c}$. While with the correction, they become $b_1 = 0.01 \pm 0.02$ and $0.16 \pm 0.03$, respectively. 
For without the star-forming gas correction case, the SO method better reproduce the {\HI} power spectrum than the pasting method. On the other hand, for without the correction case, the pasting is better than the SO method.}

Finally, our SO method has the advantage of simplicity in that only a single parameter $R_{\rm up}/R_{200c}$ for the radius is required to reproduce the {\HI} power spectrum, although with some limitation 
on the slope of the computed power spectrum with the chosen value of $R_{\rm up}/R_{200c}$.
In contrast, the pasting method highly depends on the relation between the $M_{\rm halo}-M_{\HI}$ 
which is sensitive to the resolution of the hydrodynamic simulation. Also, the relation shows large scatters and difficult to describe it with an analytic model.  \cite{TNG2018} modelled the relation with three parameters ignoring the scatter around this relation. It should be noted however that the scatter around the relation does not affect the resulting power spectrum very much \citep{Modi+:2019}. 
\RA{The {\HI} power spectrum on large scales is mainly determined by the total {\HI} mass of the halo. The pasting method assigns the {\HI} mass to each halo using the $M_{\rm halo}-M_{\HI}$ relation directly which is calibrated from the hydrodynamic simulation with or without the star-forming gas correction.}

\RAm{After comparing the goodness of the model, with the star-forming gas correction, the pasting method better reproduces the shape of the {\HI} power spectrum than the SO method. However, it goes the opposite in the case without the star-forming gas correction. In addition to the star-forming gas correction, non-trivial astrophysics effects may alter the 21-cm line power spectrum. 
This means that the reconstruction of {\HI} power spectrum is subject to the many complex astrophysical effects.
Therefore, 
considering a variety of 
reconstructing {\HI} power spectrum is essential. 
The pasting method tells us the {\HI} mass in the halos can well reproduce the {\HI} power spectrum using $M_{\rm h}-M_{\HI}$ relation.  In this paper, the SO method not only indirectly alters the mass function through the $R_{\rm up}$ but also it takes into account the spatial density distribution in and around the halo \citep[see also][]{2020arXiv201206795S}. We find those effects are also vital to reconstruct the {\HI} power spectrum.
}

\begin{table}
\caption{The best-fitting parameters for $M_{\rm h}-M_{\HI}$ relation with and without the star-forming gas correction.
\label{tab:fitting}}
\begin{tabular}[width=\columnwidth]{c|ccc} \hline\hline

                    & $\alpha$      & $M_{\rm min} [{\rm M_\odot}]$ & $ M_{0}\,[{\rm M_\odot}]$  \\ \hline
     w/ correction  & $0.82 \pm \RAm{0.02}$ & $(1.15 \pm \RAm{0.07})\times 10^9$      & $( 3.0\pm \RAm{0.2})\times 10^{10}$ \\
     w/o correction & $0.54 \pm \RAm{0.01}$ & $(1.21 \pm \RAm{0.06})\times 10^9$      & $(4.9\pm \RAm{0.3})\times 10^{10}$ \\

     \hline\hline
\end{tabular}
\end{table}

\subsection{Effect of inhomogeneous spin temperature}
\label{ssec:spintemp}
In the context of cosmological analysis, the {\HI} power spectrum is the main concern in the literature \citep{Bharadwaj+2005,Sarkar+2016,Sarkar+2018,Padmanabhan+2017a, Padmanabhan+2017b,Modi+:2019}. 
However, what we observe by radio interferometers is the intensity of radio flux or in other words, brightness temperature. \citep[e.g.][]{Madau+:1997,Ciardi+:2003}. 
As shown in Section~\ref{ssec:densityfield}, not only the fluctuations on neutral fraction and gas density but also the fluctuation on spin temperature affect to produce the brightness temperature fluctuation. 
If the temperature is high enough $T_{\rm \gamma}/T_S\ll 1$, which is usually the case within the galaxy or circum-galactic medium, the fluctuation of spin temperature can be ignored and neutral hydrogen is only a source of fluctuation of $\delta T_b$. 
However, in the case 
the gas temperature is significantly low like in the inter-galactic medium, 
the spin temperature may also not be high enough and its fluctuation  
is imprinted on
the 21-cm line power spectrum. 

\begin{figure}
\centering
	\includegraphics[width=\linewidth]{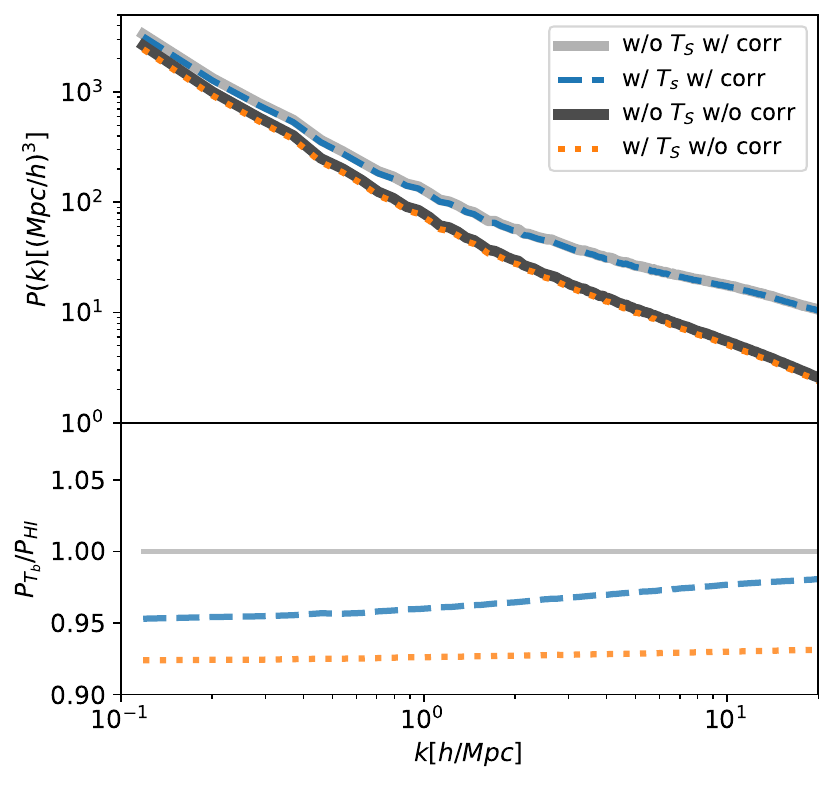}
    \caption{
    Comparing the {\HI} power spectrum  $P_{\HI}(k)$ (solid line) and the brightness temperature power spectrum $P_{T_b}(k)=P_{n_{\HI}(1-T_{\gamma}/T_S)}(k)$ without (orange dashed line) and with the star-forming gas correction (blue dashed line).  The grey line is $P_{\HI}(k)$ with the star-forming gas correction. 
    \label{fig:PkTb}}
\end{figure}

In this section, we calculate the brightness temperature power spectrum 
by considering the spin temperature fluctuation. 
\RAm{In order to factorize the effect of spin temperature fluctuation on the power spectrum, we compare the power spectrum with different two prescriptions,
\begin{align}
    \label{eq:Tb1}
    &\delta T_b \propto \overline{\rho_{\HI}}(1+\delta[\rho_{\HI}]) \\
    \label{eq:Tb2}
    &\delta T_b \propto \overline{\rho_{\HI}T_{S\gamma}}(1+\delta[\rho_{\HI}T_{S\gamma}]),
\end{align}
where $T_{S\gamma}\equiv(1-T_{\gamma}/T_S)$. In equation (\ref{eq:Tb1}), $T_{S\gamma}$ is assumed to be unity, which has been considered reasonable at the post-recombination epoch in the literature. On the other side, in equation (\ref{eq:Tb2}), the spin temperature is also fluctuating in addition to the neutral hydrogen gas density, and then we compute the total fluctuation of the brightness temperature. }
Figure~\ref{fig:PkTb} 
compares the power spectra defined by equations (\ref{eq:Tb1}) and (\ref{eq:Tb2}) apart from the normalization prefactors. 
We see that the spin temperature fluctuation causes a slope deviation of about 2\% in $k\leq1\hmpci$. 
Reminded that the power spectrum at small scales is dominated by high-density regions and large scales by lower density regions. Therefore, the effect of $T_{S\gamma}$ which largely impacts at low-density regions may suppress the amplitude on large scales.

Figure~\ref{fig:PkTb} also shows the effect of spin temperature depending on the star-forming gas correction.
Since the star-forming gas correction 
increases {\HI} mass in the high-density region, both this correction and the spin temperature fluctuation produce an effect that adds weight to the high-density region. Thus, the 
spin temperature fluctuation is more effective in the case with the star-forming gas correction. 
\subsection{Density profile in the halo}
It is of great interest to see the density profile of {\HI} within and around the halos both for the cosmological analysis and astrophysical understandings such as feedback.
In Section~\ref{ssec:dmso}, we see that the {\HI} power spectrum is reproduced well using the dark halo power spectrum when the halo radius is extended to certain scales. 
We are going to explain how this coincidence came about by shedding light to the difference of density profiles of dark matter and {\HI}.

We classify all the dark matter and gas particles into FoF or non-FoF (we call the latter as IGM particles, hereafter). The radial profile is then measured for each halo to eliminate the effect of neighbouring halos at large halo centric radii. The density profile for IGM particles is also measured separately. We then stack all the halos in the narrow mass bins: 
3000 halos for the mass ranges 
($\log_{10}(M_{\rm halo}[\himsun]) \geq$ 9, 10, 11, and 12), 
and all the halos for mass ranges ($1\times 10^{12}<M_{\rm halo}[\himsun]<2\times 10^{12}$). 

\begin{figure*}

    \centering
    
    \begin{tabular}{cc}
	\includegraphics[width=0.5\linewidth]{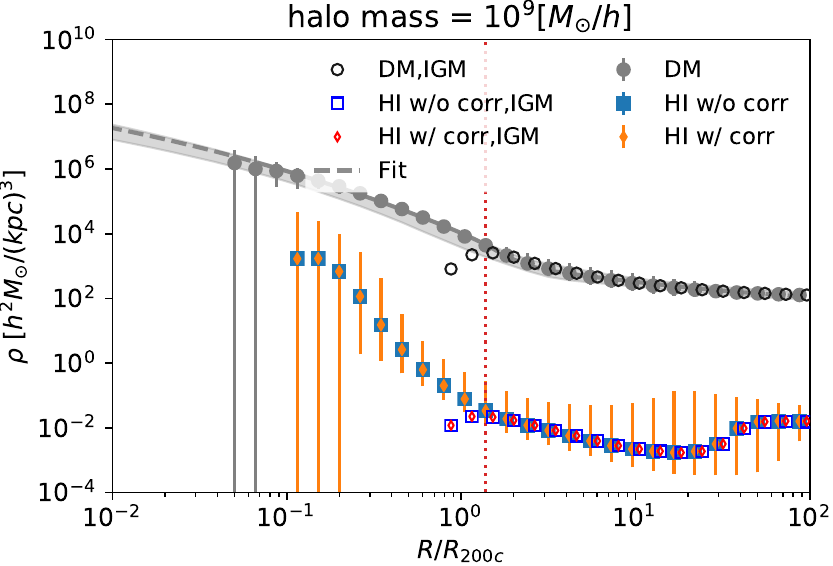} &
	\includegraphics[width=0.5\linewidth]{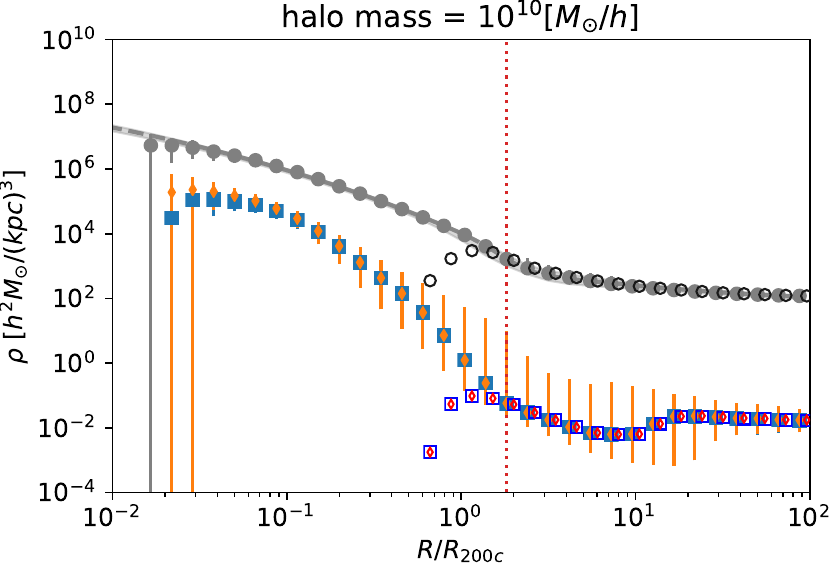} \\
	\includegraphics[width=0.5\linewidth]{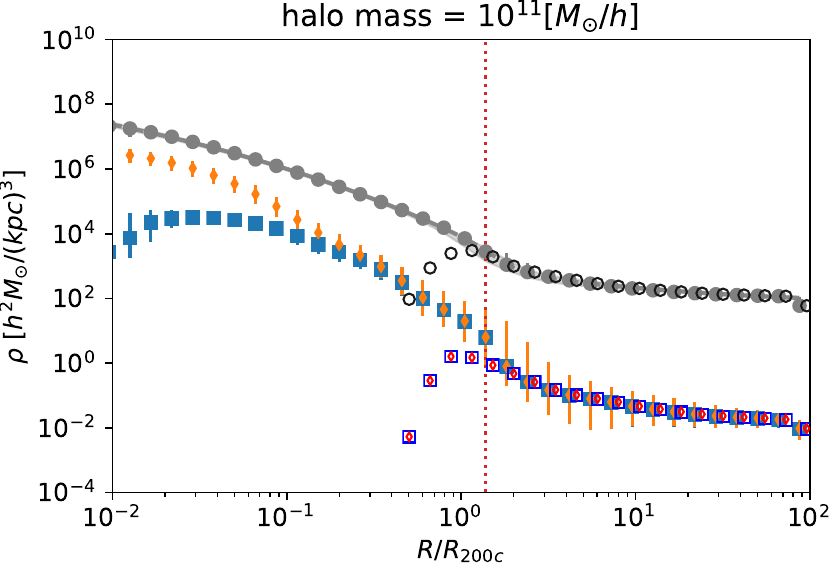} &
	\includegraphics[width=0.5\linewidth]{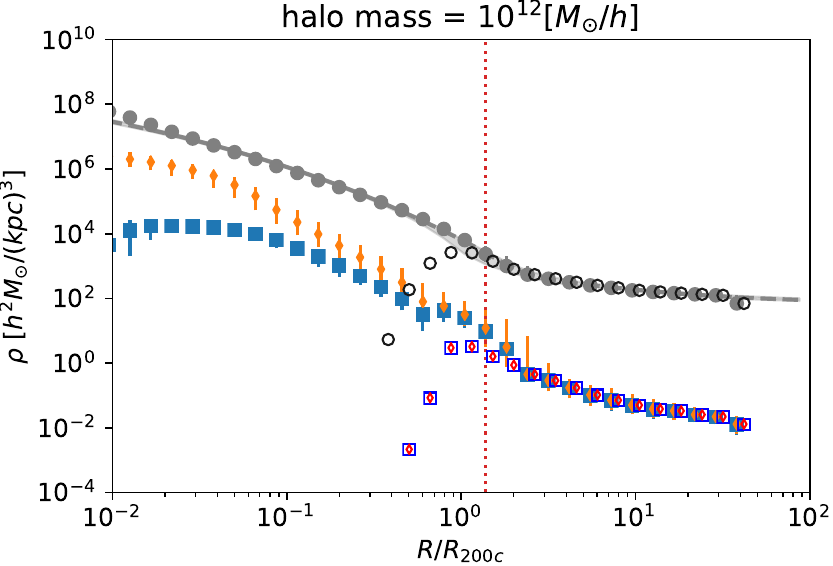} \\
    \end{tabular}
    \caption{
    Stacked {\HI} and dark matter density profiles for $\log_{10}(M_{\rm halo}[\himsun]) \sim$ 9, 10, 11, and 12 from the upper left to the bottom right panel. 
The grey dashed line shows the best-fitting model for dark matter profile, and the shaded region represents the uncertainty of the fit. 
For stacking, we use up to the 3000th halo counting from the lower end of the mass $\log_{10}(M_{\rm halo}[\himsun]) \geq $ 9, 10 and 11. However for $\log_{10}(M_{\rm halo}[\himsun]) =12$, we use all halos with masses of $1\times 10^{12}<M_{\rm halo}[\himsun]<2\times 10^{12}$. 
The vertical dotted red line represents the splashback radius. 
The open symbols are the profiles measured using only the IGM particles, and the closed ones are those using the FoF and IGM particles. 
The open symbols are slightly shifted to the right for a better visibility. 
	    \label{fig:HIDMrho}}
\end{figure*}

Figure~\ref{fig:HIDMrho} shows the dark matter and {\HI} profiles around the halos for four different halo masses. For a wide range of mass of halos except for the very massive ones, 
the {\HI} density profile is steeper than that of dark matter and thus the {\HI} is relatively more concentrated on the halo centre and dark matter has more diffuse distribution. For the purpose of reproducing the {\HI} power spectrum from dark matter, the simplest way is to remove the diffuse dark matter outside the halo as we discussed in Section~\ref{ssec:dmso}, 
and we took the truncation radii $R_{\rm up}=2R_{200c}$ and $0.6R_{200c}$ for with and without star-formation correction, respectively. 
Now the question is how this truncation scale is determined?  

To partly answer this question, we first compare the results with and without star-formation correction. With the star formation correction, the {\HI} is more concentrated near the centre of halos. This is because the correction puts an upper limit on the gas temperature at $10^4$K which may diminish the ionization of star-forming gas located at the halo centre region, and thus increase the abundance of the {\HI}. This implies that with the star-formation correction, the diffuse dark matter should be truncated at smaller scales compared to the case of without correction. In fact, we have observed that the truncation scale is three times smaller for the case of the star-formation correction.

Another aspect to answer the aforementioned question is to find a physically motivated scale of the truncation. Recent high-resolution N-body simulations have revealed that there is a clear boundary around the dark matter halo, the so-called splashback radius. 
Around the splashback radius, the accreting mass reaches the apocentre and is piled up due to low radial velocity \citep{Fillmore+1984,Bertschinger1985,Adhikari+2014,Diemer+2014,Mansfield+2017}.
In other words, there is a rapid decline in the density just outside the splashback radius, and there is a density gap. 
Since the probability of finding mass outside the splashback radius is significantly reduced, it seems reasonable to define the truncation scale to be the splashback radius. In practice, the splashback radius is defined as the radius where the logarithmic derivative of the density profile becomes minimum, the steepest point. 

To measure the splashback radius, we fit the stacked profiles 
with the following density jump model 
\citep{Diemer+2014}:
\begin{align}
    \rho(r) &= f_{\rm trans}\rho_{\rm Einasto}+\rho_{\rm outer}, \notag\\
    \rho_{\rm Einasto} &= \rho_s\exp\left(-\frac{2}{\alpha}\left[\left(\frac{r}{r_s}\right)^{\alpha}-1\right]\right), \notag\\
    f_{\rm trans}&=\left[1+\left(\frac{r}{r_t}\right)^{\beta}\right]^{-\gamma/\beta}, \notag\\
    \rho_{\rm outer} &= \rho_m\left[b_e\left(\frac{r}{5R_{\rm 200c}}\right)^{-S_e}+1\right].
    \label{eq:rhomodel}
\end{align}
\RAm{We note that in \citet{Diemer+2014}, the outer profile is scaled with $R_{\rm 200m}$, which is the radius where the mean density inside is 200 times the mean density of the Universe. 
We replace it with $R_{200c}$ which is used consistently throughout this paper, and the difference between them are negligibly small in particular at high redshifts. }
We fix the parameters as $\beta=6$ and $\gamma=4$ as in \citet{Diemer+2014, More+2015}, and fit the other five parameters within the range of $0.1R_{200c}$ to $10R_{200c}$ to the dark matter profiles. Once we find the best-fitting parameters, we differentiate the model to find the splashback radius, $R_{\rm sp}$. 
For most of the halo masses, the splashback radius is about $1.35 R_{200c}$ and slightly smaller at the massive end, $M\geq 10^{13} \himsun$, consistently with \cite{More+2015}. 
We denote the ratio as $C(M)\equiv R_{\rm sp}/R_{200c}$. 
Finally, we make a density map with the mass-dependent SO radius and $C(M)$, and measured the power spectrum. 
\RAm{The scale-dependent terms of the ratio $\sqrt{P_{\HI}/P_{\rm so}}$ are $b_1=0.03\pm 0.03$ and $b_1 = 0.27\pm 0.03$ for the case without and with the star-forming gas correction, respectively. }
For the case without star-forming gas correction, this $b_1$ value is consistent with zero, and larger than that for the constant SO-radius case ($b_1=-0.005$ at $R_{\rm up}/R_{\rm 200c}=2$). 
Since the smaller value of $b_1$ means better reproduction of the scale dependence of the {\HI} power spectrum on large scales, the splashback radius is 
a reasonable quantity to use for the case with the star-forming gas correction.

\begin{figure*}
    \centering
    \begin{tabular}{cc}

	\includegraphics[width=0.45\linewidth]{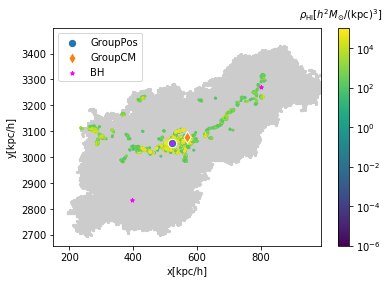}&
	
 	\includegraphics[width=0.45\linewidth]{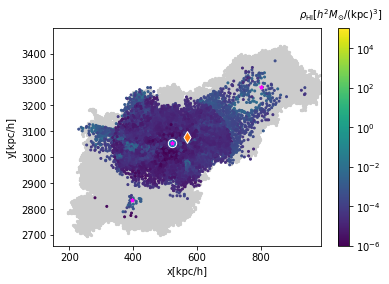}	

	    \end{tabular}
    \caption{
    {\it Left panel:} Density distribution of the dark matter (grey points) and \HI\ gas (coloured points) of a certain FoF-halo with a mass of $1.2\times 10^{12}\himsun$. 
    This figure only shows the {\HI} with high densities of $\rho_{\HI}\geq 10^2 h^2\Msun {\rm kpc}^{-3}$. The colour gradient shows the value of the \HI\ density. The positions of the halo centre, \texttt{GroupPos} (blue circle), \texttt{GroupCM} (orange diamond), and the black hole (magenta stars) are also plotted.
    {\it Right panel:} Same as the left panel but for the lower density \HI\ gas with $\rho_{\HI}\leq 10^{-2} h^2\Msun {\rm kpc}^{-3}$.
	    \label{fig:HIdenshalo}}
\end{figure*}
In order to 
illuminate the individual {\HI} distribution within the halo, we show the 
particles in the FoF group in Fig.~\ref{fig:HIdenshalo}. 
The left panel shows the distribution of high density {\HI} ($\geq 10^{2} h^2\Msun {\rm kpc}^{-3}$),
and the right panel is for the lower \HI\ density ($< 10^{-2} h^2\Msun {\rm kpc}^{-3}$).
We also plot the position of the halo centre and BH in both panels. 
As it can be seen from this figure, there is an under-dense region around the BH like a bubble 
due to AGN feedback. At the same time, we can see 
the cold flow, a filamentary structure of high density \HI\ around the BH. 
The halo centre, represented by \texttt{GroupPos}, overlaps with the position of the BH. 
The radial density profile in Fig.~\ref{fig:HIDMrho} shows that there is a slightly low \HI\ density region around the halo centre for the case without the star-formation correction. In other words, 
the peak of the {\HI} density and the position of the AGN are apart. 
The recent study about the cross-correlation between Lyman-$\alpha$ absorption and AGNs \citep{Mukae+2020,Momose+2020} also show that there is an under-dense region due to photoionization by the AGN feedback, and the peak of the {\HI} density is a few Mpc away from the position of the AGN.
Besides, they also suggest that {\HI} density around the AGN depends on the environments and the AGN type. 
Hence, the density profile of the {\HI} depends on the astrophysics, and the model of the {\HI} density profile can be used in the modelling of 21-cm line survey and Lyman-$\alpha$ tomography. 

\subsection{Scale dependence at different redshifts}
\label{ssec:assessing}
\begin{figure}
    \centering
    \includegraphics[width=\linewidth]{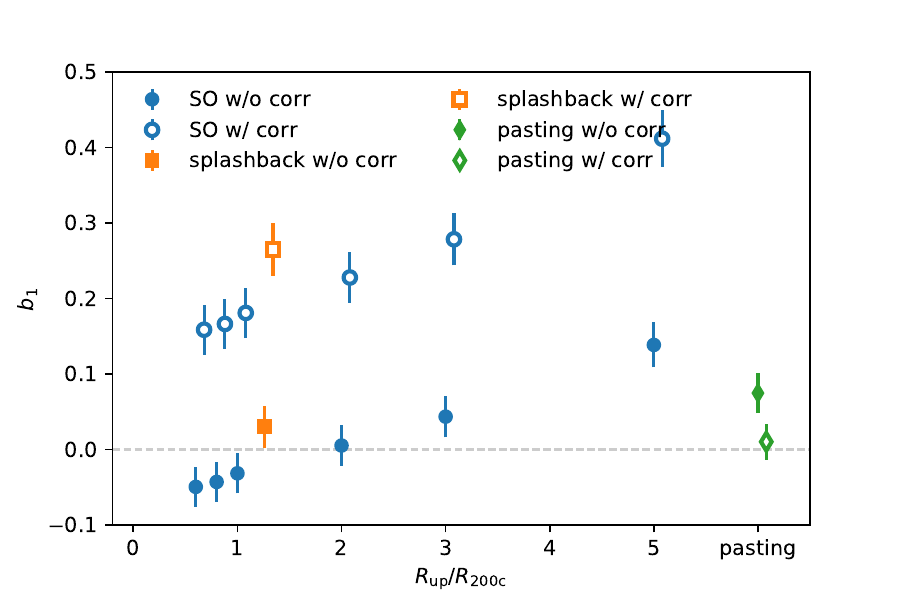}
    \caption{
Best-fitting parameters of the scale-dependent term $b_1$ for $k<1\hmpci$ in Eq.\,\ref{eq:kfitfunc} 
for the cases using the SO radius, pasting, and splashback radius. 
The horizontal line of $b_1 = 0$ means that there is no scale-dependent bias. 
	    \label{fig:Psp}}
\end{figure}

\begin{figure}
    \centering
    \includegraphics[width=\linewidth]{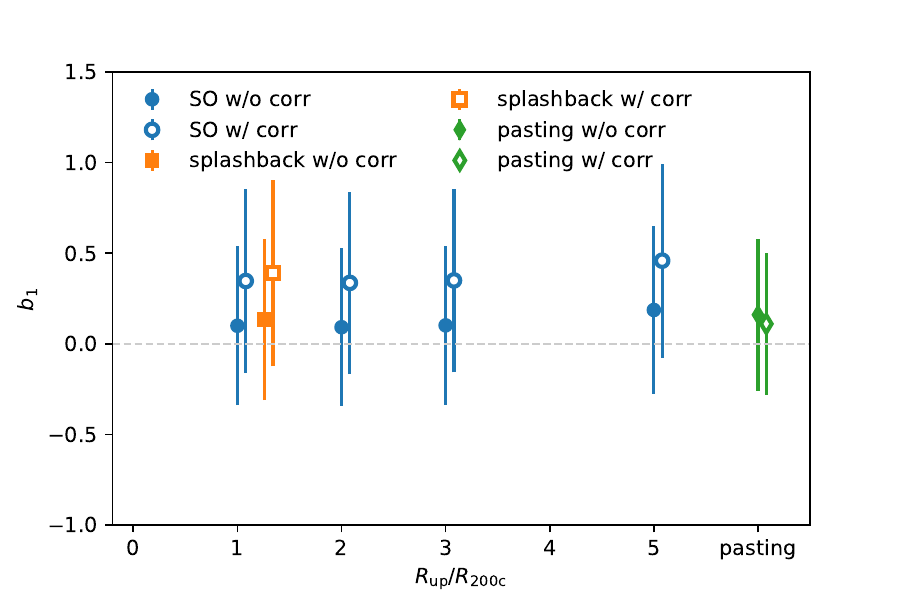}
    \caption{
Best-fitting parameters of the scale-dependent term $b_1$ for the first three points at $k<0.3\hmpci$.
	    \label{fig:Psp_kmax03}}
\end{figure}

\RAm{In this section, we evaluate each model for reconstruction by comparing the scale dependence of the ratio of the reconstructed power spectrum and {\HI} power spectrum. 
As introduced in section \ref{ssec:pasting}, we parameterize the model residual as $\sqrt{P_{\HI}/P_{\rm Model}}=b_0+b_1k$ in order to evaluate the scale dependence.
Figure~\ref{fig:Psp} summarises the best-fitting $b_1$ values 
for different methods obtained at 
$k\leq 1\hmpci$. }
\RAm{
For $M_{\rm halo}-M_{\HI}$ pasting, $b_1$ is consistent with zero 
for the case of star-forming gas correction.
For the SO method, $b_1$ at $R_{\rm up}=2R_{200c}$ is 
consistent with zero for the case without the gas correction. Conversely, $b_1$ values at other SO radii are away from zero. 
This is partly due to the larger $k_{\rm max}$ for the fitting where the most of the detectable BAO wiggles are well located at k< 1$\hmpci$.
If we focus on the first few peaks and troughs of the BAO wiggles, 
we only need the {\HI} power spectrum on much larger scales. 
Figure~\ref{fig:Psp_kmax03} is 
same
as Fig.~\ref{fig:Psp} but using only 
scales $k\leq 0.3\hmpci$ for fitting the residual. We see that for all cases, $b_1$ values are consistent with zero within 1-sigma error both for the case with and without the star-forming gas correction, although the error is large. 
}

\begin{figure}
    \centering
    \includegraphics[width=\linewidth]{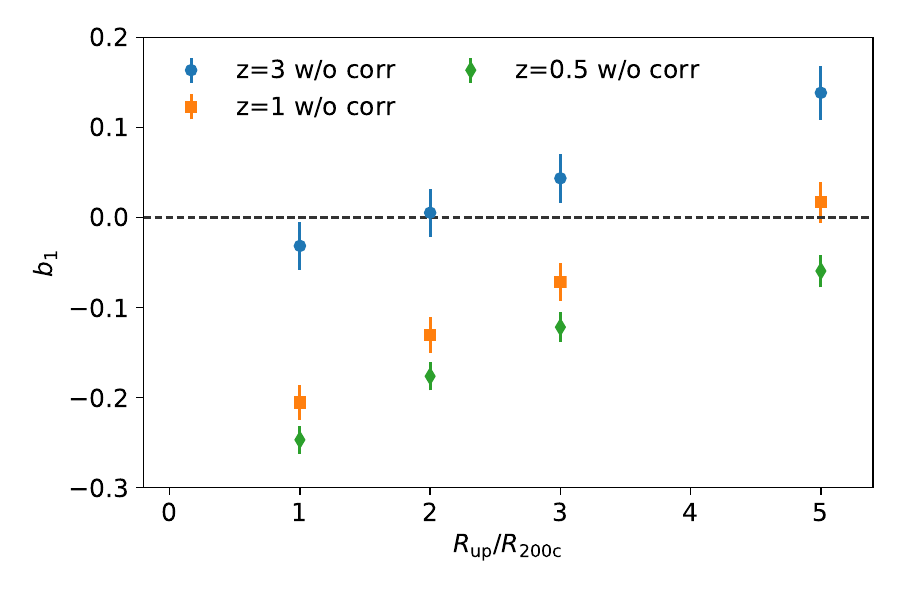}
    \caption{
\RAm{Best-fitting parameters of the scale-dependent term $b_1$ for $k<1\hmpci$. Green diamond, orange square, and blue circle are for $z=0.5, 1, 3$, respectively. Open symbols show the $b_1$ with the star-forming gas correction, and closed symbols show that without the correction.} 
	    \label{fig:Psp_allz}}
\end{figure}

\begin{figure}
    \centering
    \includegraphics[width=\linewidth]{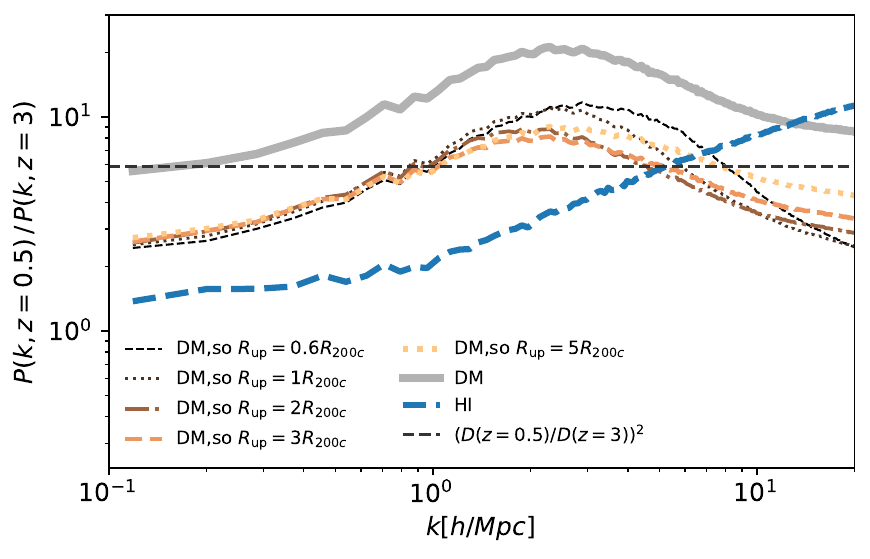}
    \caption{
\RAm{The ratio of the power spectrum at $z=0.5$ and $z=3$. 
Gray solid line is for the dark matter power spectrum, and blue dashed line is for the {\HI} without the star-forming gas correction. Other lines are the DM,so with different $R_{\rm up}$. The horizontal dashed line represents the linear growth rate from $z = 3$ to $z = 0.5$.} 
	    \label{fig:Pk_growth}}
\end{figure}

\RAm{We also investigate the redshift dependence of the best $R_{\rm up}$.  Figure~\ref{fig:Psp_allz} shows the $b_1$ values defined in equation \ref{eq:kfitfunc}
at $z=0.5, 1,$ and $3$. We see that the $b_1$ monotonically increases from $z=0.5$ to $3$, and thus the best $R_{\rm up}$ decreases. The increase of $b_1$ is naturally explained by the difference in the way of non-linear evolution of the power spectra between DM,so and {\HI}.} 
\RAm{Figure~\ref{fig:Pk_growth} shows the growth of the power spectrum from $z=3$ to $z=0.5$. We see that the power spectrum of the dark matter grows at a \RAm{rate of the linear growth factor}, $(D(z=0.5)/D(z=3))^2$ on large scales. 
\RAm{On small scales, the dark matter power spectrum grows more strongly 
due to the 
non-linear gravitational growth of the structure. 
The growth of the DM,so power spectrum}
 is smaller than that of the linear growth, 
\RAm{because it is suppressed by the halo bias evolution, given that the halo is a less-biased tracer at low redshifts. Interestingly the non-linear boost of the DM,so is similar to that of the dark matter power spectrum, presumably because the dark matter still exist in the IGM and will be accreted onto halos.} We also see \RAm{that} the ratio of the {\HI} power spectrum 
is even smaller than that of DM,so for $k<1\himpc$. This is because most of the {\HI} is already \RAm{confined within} the halo \RAm{at $z=3$} and the {\HI} is almost absent in the IGM at $z<3$, so that the growth of the {\HI} clustering is \RAm{much} smaller than \RAm{that of} the DM,so power spectrum. 
\RAm{As a consequence, on small scales,}
the evolution of the {\HI} power spectrum is 
\RAm{weaker than that of the DM,so. Therefore, the residual function, ${\rm Res}(k)$ has more scale dependence at lower redshifts, which results in negatively large value of $b_1$ in Fig.~\ref{fig:Psp_allz}.}
}

\RAm{
In this paper we propose to model the {\HI} power spectrum from the dark matter density fields on large scales. 
Due to the limitation of the small volume of the hydrodynamic simulation, we are unable to measure the BAO scales and quantify the systematic effect due to the redshift space distortion.
In our future analysis, we will measure the BAO scales with an accuracy of 1\%, in which case, $b_1$ has to be consistent with zero with $\sigma_{b_1}<0.01$. In order to meet this requirement, we need a larger simulation box size, $L_{\rm box}=500 \himpc$
}

\section{Summary}
\label{sec:summary}
We propose a new method for reconstructing {\HI} power spectrum from N-body simulation for the purpose of 
cosmological analysis assuming future 21-cm line intensity mapping surveys.
At the epoch of post-reionization, hydrogen in the IGM is ionized by the UV background, and thus most of the {\HI} is in the high-density region such as the dark matter halo. 
Therefore, some previous studies assign the {\HI} at the centre of the halo assuming the {\HI} mass and halo mass relation to reproduce the {\HI} power spectrum \citep{Sarkar+2016,Sarkar+2018,Modi+:2019,Wang+:2019,TNG2018}. 

In this paper, we first see the effect of {\HI} in the IGM on the power spectrum using IllustrisTNG simulation.
We find that not only within the halo, but also the {\HI} around the halo at radii $r<3R_{\rm 200c}$ contributes to the {\HI} power spectrum. Therefore, {\em it is important to model the {\HI} distribution in and around the halo, and not just by using the relation to the halo mass.}  

Then, we propose a new method for reconstructing {\HI} power spectrum and test it. 
Our method to express the {\HI} is simply truncating the dark matter distribution at certain halo-centric radius. The truncation scale $R_{\rm up}$ is determined by monitoring the reconstructed power spectrum and real {\HI} power spectrum. 
As a result, we find that, apart from the amplitude, $R_{\rm up}=2R_{\rm 200c}$ best reproduces the slope of the {\HI} power spectrum. 

In addition, we discuss the effect of the correction for the star-forming gas 
on the {\HI} power spectrum \citep{TNG2018}. 
The correction is motivated by the fact that the IllustrisTNG simulation underestimates the total amount of {\HI} by a factor of two, compared to the DLA observations \citep[e.g.][]{Rao+2006, Lah+2007,Noterdaeme+2012,Songaila+2010,Crighton+2015}. 
Although considering the molecular hydrogen slightly decreases the {\HI} mass, the overall {\HI} mass increases when the temperature floor of $T=10^4$\,K is imposed on the star-forming gas.  
This correction increases the {\HI} mass, especially in massive halos. 
As a result, the amplitude of the 1-halo term of the {\HI} power spectrum becomes higher, 
and the damping scale of the 2-halo term moves to a larger scale. 
Therefore, the {\HI} power spectrum with the correction is flatter than that without the correction. 
With this correction, the best choice is $R_{\rm up}/R_{\rm 200c}=0.6$ to reproduce the {\HI} power spectrum, but it does not fully reproduce the scale dependence of the {\HI} power spectrum, 
and the agreement with the true {\HI} power spectrum is worse than the case without the star-forming gas correction.
We argue that it would be reasonable to make a connection between the truncation scale and the `splashback' radius \citep{Diemer+2014},  
which roughly corresponds to $1.3R_{200c}$ but slightly smaller for more massive halos. 
We have shown in this paper that it can also provide a good representation of {\HI} power spectrum. 

Furthermore, we compare these results with a previous method which pastes the {\HI} on the halo centre with $M_{\HI}-M_{\rm halo}$ relation to see how well they reproduce the scale dependence of the {\HI} power spectrum on large scales, $1\leq \hmpci$, for the measurements of the BAO scales. 
The SO method can reproduce the {\HI} power spectrum using only a single parameter, but it is limited in the shape of the power spectrum that can be reproduced. 
On the other hand, the pasting method requires several parameters for the $M_{\HI}-M_{\rm halo}$ relation, but it can reproduce the shape of {\HI} power spectrum well, irrespective of the star-forming gas correction. 

So far, we only consider the {\HI} inhomogeneity assuming that the spin temperature is spatially uniform at the post-EoR. It is normally true in the region where the gas density and temperature is high. However, it is not necessarily correct in the low-density region such as cosmic voids. We explore the effect of inhomogeneous spin temperature on the power spectrum of brightness temperature, which is the real observable in actual observations. We find that the effect of inhomogeneous spin temperature is imprinted on the power spectrum and suppresses the amplitude by 5\% and 8\% with and without the star-forming gas correction, respectively. 
However, we see negligible changes in the slope of the power spectrum. 

In the future 21cm intensity mapping survey such as SKA, we expect to determine the BAO scale with an accuracy better than 1\%. In order to examine whether our model can reproduce the {\HI} power spectrum with this accuracy, a larger simulation box size of 500 Mpc/$h$ will be required. However, given the limited computational resources, we cannot achieve the same mass resolution in such a large box size simulation at this moment. This issue will be studied in our future work.

\section*{Acknowledgments}
We would like to thank Kenji Hasegawa, Arman Shafieloo and Khee-Gan Lee for useful discussions. 
This work is supported by Grant-in-Aid for JSPS Fellows JP19J23680 (RA) and the JSPS KAKENHI Grant Number
15H05890, 20H01932 (AN) and JP17H01111, 19H05810 (KN). 
KN acknowledges the travel support from the Kavli IPMU, World Premier Research centre Initiative (WPI). 

\section*{Data availability}
The data underlying this article are available in IllustrisTNG simulation, at https://www.tng-project.org. 

\bibliographystyle{mnras}
\bibliography{bibdata}

\bsp	
\label{lastpage}
\end{document}